\documentclass[lettersize,journal]{IEEEtran}
\usepackage{amsmath,amsfonts}
\usepackage{algorithmic}
\usepackage{algorithm}
\usepackage{array}
\usepackage[caption=false,font=normalsize,labelfont=sf,textfont=sf]{subfig}
\usepackage{textcomp}
\usepackage{stfloats}
\usepackage{url}
\usepackage{verbatim}
\usepackage{graphicx}
\usepackage{cite}
\usepackage{amssymb}
\usepackage{multirow}
\usepackage{booktabs}
\usepackage{bm}
\usepackage[dvipsnames]{xcolor}
\usepackage{tikz}
\usepackage{placeins}
\begin{document}

\title{Uncertainty-driven Sampling for Efficient Pairwise Comparison Subjective Assessment}

\author{Shima Mohammadi, \IEEEmembership{Member, IEEE}, Jo\~{a}o Ascenso,  \IEEEmembership{Senior Member, IEEE}
\thanks{The authors are with the Instituto Superior Tecnico and Instituto de Telecomunicacoes, 1049-001 Lisboa, Portugal (e-mail: shima.mohammadi@lx.it.pt; joao.ascenso@lx.it.pt).}
}



\maketitle

\begin{abstract}
Assessing image quality is crucial in image processing tasks such as compression, super-resolution, and denoising. While subjective assessments involving human evaluators provide the most accurate quality scores, they are impractical for large-scale or continuous evaluations due to their high cost and time requirements. Pairwise comparison subjective assessment tests, which rank image pairs instead of assigning scores, offer more reliability and accuracy but require numerous comparisons, leading to high costs. Although objective quality metrics are more efficient, they lack the precision of subjective tests, which are essential for benchmarking and training learning-based quality metrics. This paper proposes an uncertainty-based sampling method to optimize the pairwise comparison subjective assessment process. By utilizing deep learning models to estimate human preferences and identify pairs that need human labeling, the approach reduces the number of required comparisons while maintaining high accuracy. The key contributions include modeling uncertainty for accurate preference predictions and for pairwise sampling. The experimental results demonstrate superior performance of the proposed approach compared to traditional active sampling methods. An implementation of the pairwise sampling method is publicly available at \url{https://github.com/shimamohammadi/LBPS-EIC}
\end{abstract}


\section{Introduction}
\IEEEPARstart{I}{mage} quality evaluation plays a vital role by quantifying quality degradations according to human perception, which is especially required in many visual processing tasks, such as compression, super-resolution, denoising and others. There are two primary methods for evaluating image quality: subjective assessment tests and objective quality metrics. Subjective assessment, considered the most accurate, involves gathering quality scores from human evaluators using a reliable methodology and graphical interface. However, despite its precision, this approach can be impractical in many scenarios, e.g., during the optimization or training of image compression algorithms or for continuous quality monitoring. Consequently, objective metrics, which automate the quality assessment process, are frequently used. 
\par Depending on number of stimuli that are present to the subjects, several subjective assessment methodologies can be used to obtain quality scores, including single stimulus and double stimulus approaches. Single and double stimulus methodologies like Absolute Category Rating (ACR) and Double Stimulus Impairment Scale (DSIS) are even standardized\cite{ITU-P-913, BT500-19}. These methodologies require a panel of human subjects to score each visual stimulus on a predefined category scale. The average of these scores is then reported as mean opinion scores (MOS) or differential MOS (DMOS) relative to a reference. However, despite the widespread use, it lack sensitivity when the quality differences between stimuli are small and also suffer from different types of biases, past experiences, or expectations. For example, the interpretation of the category scales can vary among subjects, the perception of ``Excellent'' quality might differ among subjects in different contexts, or, they may change their ratings during the test, leading to inconsistent scores. These limitations can result in less reliable test outcomes.

\par A promising alternative to address the aforementioned issues is to simplify the subjective evaluation process by replacing category rating with simple ranking. Ranking approaches compare image pairs or triplets to determine which one is of higher quality. Typically, it is much easier (and leads to an increase sensitivity) for human subjects to express a preference between two images rather than attribute a score for each. Pairwise comparison (PC) is an example of a ranking methodology, where subjects are asked to select which image of a pair they find more preferable. The preferences can be for all possible pairs and quality scores can be inferred using models such as Thurstone \cite{TM} or Bradley-Terry\cite{Bradley1952}. This methodology leads to more accurate and reliable results than category rating methodologies \cite{Pinson2003}\cite{Tominaga2010}\cite{Mantiuk2012}. 
In such case, subjects in pairwise comparison tests are better able to detect small quality differences and are less affected by changes in viewing conditions. This approach is becoming rather popular nowadays for assessing high-quality to nearly visually lossless images and is the primary focus of the JPEG AIC ad-hoc group which is dedicated to standardizing both subjective and objective quality metrics for image compression.

\par Although the pairwise comparison test is rather accurate, it may require a large number of pairs, making it costly in most cases, especially because to assess the quality of $n$ distorted images, $n(n-1)/2$ pairs are needed if all possible pairs are evaluated (all images are compared to each other). In this case, the number of pairs grows quadratically $\mathcal{O}(n^2)$, with the number of distorted images. Consequently, when using many reference images and their distorted versions, the pairwise comparison test becomes impractical or very expensive. One solution to this problem is to sample only a subset of pairs for the pairwise comparison test. A straightforward approach is to restrict some types of pairs (e.g. between different types of degradation) or to randomly select a set of pairs for evaluation. However, several pairwise sampling methods were already proposed. Nowadays, the most promising approach is to select pairs based on past decisions made by subjects and according to a measure of the impact of the pair on the final scores. However, this requires running the sampling procedure after each comparison (trial) for every pair, making it impractical in many scenarios. Moreover, this sampling relies on past decisions and can be compromised by malicious subjects, leading to inefficient sampling. Therefore, it is preferable to sample a subset of pairs beforehand, independent of the subjective test results. Therefore, the key idea in this paper is significantly different from previous state-of-the-art pairwise sampling methods: some of the ranking decisions can be automatically predicted using a deep learning framework, while human subjects will make the remaining decisions.

\par In this context, the objective of this paper is to propose uncertainty-based sampling to reduce the subjective test (following a pairwise comparison procedure) duration without compromising the accuracy of the results. The proposed uncertainty models will estimate human preferences between image pairs and identify which pairs require active human labeling. Unlike previous work, this approach will be implemented using a deep learning model for the first time. The novel concept is that pairs with unreliable preference estimations will be deferred to human subjects for accurate preferences, while the estimated preferences will be directly used for the remaining pairs. 

\par Therefore, the main contributions of this paper are summarized below.
\begin{itemize}[leftmargin=*]
    \item \textbf{Modeling uncertainty for accurate preference prediction}: Propose a data (or aleatoric) uncertainty estimator to obtain an accurate prediction of the probability of preference within a pairwise comparison methodology. By leveraging deep learning models, the goal is to obtain reliable estimates of the underlying quality of both images in a pair, thereby modeling human subjects’ preferences. The final estimates are used as replacement of the subjective scores for some of the pairs reducing the length of the subjective test.
    \item \textbf{Modeling uncertainty for accurate pairwise sampling}: Propose a model (or epistemic) uncertainty procedure to identify instances (pairs) where the deep learning model cannot reliably  estimate the probability of preference. In such cases, the pair of images must be subjectively evaluated. This contribution allows to select the pairs of images for which the deep learning model's predictions are uncertain and request, for these cases, the help from human subjects. 
\end{itemize}

\par The rest of the paper is organized as follows: Related work is described in Section \ref{related-work-section}. The uncertainty models are described in Section \ref{Uncertainty-section}. The quality estimation and pairwise sampling procedure is detailed in Section \ref{seq-pairwise-sampling-procedure}. In Section \ref{performance-section}, the performance evaluation including the training procedure and experimental results are described. Finally, the conclusion is provided in Section \ref{sec-conclusion}.

\section{Related Work} \label{related-work-section}
This section provide an overview of relevant related work considering the objectives of this paper, first, pairwise sampling methods are reviewed followed by uncertainty estimation methods in objective quality metrics.

\subsection{Pairwise Sampling}
To address the challenging issue of pairwise sampling on subjective tests, several techniques have been proposed in the literature. These approaches can be classified into several types: 1) random selection: involves randomly selecting a subset of pairs \cite{Xu2012a} for human evaluation, 2) sorting-based selection: treats pairwise comparisons as a sorting problem, where each subject attempts to sort the visual stimuli \cite{ponomarenko2015}, \cite{ASAP}, 3) active-based selection: iterative process where selected pair(s) are evaluated one by one or in batches by one or more subjects, their preferences are recorded and used to determine the next pairs for evaluation in the next iteration, 4) predictive sampling: involves performing a subjective test with a pre-selected short list of pairs. For the remaining pairs, the pairwise preference is automatically predicted.

\par Xu et al. in \cite{HrActive} proposes an active sampling approach based on information maximization to improve the sampling efficiency in the context of a crowd-sourcing scenario. The framework of HodgeRank (HR) is exploited: HodgeRank is a method used to decompose ranking data into consistent and inconsistent parts, providing a structured way to handle noisy and conflicting input. A supervised sampling method involving the labels already collected is proposed based on Bayesian information maximization which selects pairs with the largest information gain. Chen et al. in \cite{Crowd-BT} proposes the Crowd-BT algorithm, which extends the Bradley-Terry (BT) model to include annotator quality in crowd-sourcing. An active learning strategy is proposed that involves learning annotator quality and pairwise preferences. The experiments show that modeling annotator quality improves ranking accuracy and reduces labeling costs. Li et al. in \cite{Hybrid-MST} introduce a hybrid sampling strategy either based on minimum spanning tree or global maximum. It uses a Bayesian optimization framework along with the BT model to create a utility function, which is then used to determine the expected information gain for each pair. Mikhailiuk et al. in \cite{ASAP} propose a technique based on information gain maximization to select the most informative pairs for comparison. The algorithm updates the full posterior distribution at each iteration, offering improved performance over existing methods. Several techniques to reduce computational complexity were proposed, namely by computing an approximate (online) posterior estimation (ASAP-approx), selective expected information gain (EIG) evaluations and a Minimum Spanning Tree (MST) mode for Batch Mode. Fan et al. in \cite{ActiveSamplingTMM} combines the reliability with informativeness into an active sampling framework. Informativeness is computed based on the uncertainty or the amount of information an image pair can provide while reliability is estimated to account for the human visual system’s limitations, especially when the quality difference between two images is too subtle. 

\par Another type of methods follow a predictive sampling approach, where some of the pairs are selected, prior to the pairwise comparison subjective test. Our previous work \cite{PS-PC} proposes to train a binary classifier with a support vector machine (SVM) to determine which pairs should be included in the subjective test and a predictor with support vector regression (SVR) to estimate the preference between the images of a pair. This solution uses as features, the scores of seven popular objective quality metrics and includes a labeling procedure to obtain the ground-truth data that is essential to train the classifier. The proposed solution extends this work, by exploiting the power of deep neural networks using an approach that estimates uncertainty in both the data and the model. Unlike PS-PC, the proposed method is unsupervised and does not require a classifier, making it fundamentally different in terms of high-level system architecture.. Aldahdooh et al in \cite{Aldahdooh} proposes a method to reduce subjective testing efforts (in this case using an ACR methodology) using a smaller, representative subset of test sequences from a large dataset without compromising results. In \cite{Aldahdooh}, objective video quality metrics (e.g., PSNR, SSIM, MS-SSIM, VIF) are used to evaluate video sequences and machine learning clustering techniques such as Kmeans++, Hierarchical Clustering, and Gaussian Mixture Models (GMM) are applied to identify redundancies and similarities. This method allows significant reductions in the test size while preserving the accuracy of conclusions, however it cannot be applied for pairwise comparison methodologies.

\par The solution proposed in this paper uses a different approach since:
\begin{itemize}[leftmargin=*]
    \item The proposed method selects image pairs offline, prior to the subjective test, which differentiates it from most existing approaches — except for PS-PC — that select pairs during the test itself. This approach eliminates the need for real-time computations, unlike methods such as Hybrid-MST, which generate pairs iteratively (either one at a time or in batches), requiring continuous execution of the algorithm throughout the test. This iterative process introduces substantial computational overhead and limits the practicality and scalability of these methods, particularly in resource-limited environments. In conclusion, while our method delivers performance on par with or even surpassing that of existing approaches in terms of correlation metrics like PLCC, SROCC, and RMSE, its key advantage lies in its offline operation, boosting both efficiency and scalability, and making it especially well-suited for large-scale subjective evaluations.
    \item Different from \cite{PS-PC} instead of handcrafted feature extraction, image pair labeling and explicit training of a classifier and predictor, it is used a deep neural network with explicit uncertainty models. This new solution proposed here avoids the data labelling approach of \cite{PS-PC}, which does not consider the reliability (or confidence) in the prediction of the pairwise probabilities.
\end{itemize}

\subsection{Objective Quality Metrics}

Objective image quality metrics that exploit pairwise rankings during training are reviewed next due to its relevance with the proposed pairwise sampling procedure. Zhang et al. in \cite{UNIQUE} proposed a deep neural network model for blind image quality assessment. The new model is called UNIQUE and is trained on multiple image quality assessment databases simultaneously by computing the probability of preference between random pairs in each database; it is capable of generalizing to different types of distortions. Moreover, UNIQUE is evaluated against existing blind and full-reference image quality assessment models, demonstrating its superior performance. Cao et al. in \cite{cao2024image} design a computational framework that integrates data-centric (reliable human-annotated datasets) and model-centric (reliable quality metrics) approaches. Specifically, they propose a sampling-worthiness model that quantifies the difficulty and diversity of candidate images, identifying diverse and challenging samples for dataset construction. Ma et al. in \cite{dipIQ} address the challenge of limited training data by automatically generating quality-discriminable image pairs (DIPs) from large-scale databases, reducing the need for subjective testing. They employ RankNet, a pairwise learning-to-rank algorithm, to learn an opinion-unaware BIQA (OU-BIQA) model from these DIPs, incorporating perceptual uncertainty into the training process. The uncertainty is measured for each candidate pair using objective quality scores from three quality metrics. Liu et al. in \cite{liu2019} propose using ranking as a self-supervised auxiliary task to address the scarcity of labeled data in computer vision tasks, particularly for problems such as image quality assessment and crowd counting. In image quality assessment, they leverage on unlabeled data for training by generating distortions (adding noise, compression, etc) to reference images over a wide range of intensities; in such case, images can be ranked according to their quality since increasing distortion parameter implies decreasing image quality. Also, the network’s certainty is evaluated by performing the self-supervised proxy task (ranking) multiple times. The certainty score reflects how confident the network is in its ranking predictions which is exploited during the training process.

\par The approach followed in this work is different from the reviewed objective quality metrics (mostly with blind image quality models), due to the exploitation of novel uncertainty models for the estimation of the probability of preference in a pairwise sampling subjective test and for the corresponding reliability of this estimation. These models are of upmost importance to reduce the number of required pairwise comparisons without compromising the accuracy of the global ranking of images.

\section{Uncertainty Estimation}\label{Uncertainty-section}

\par The main idea of this paper is to exploit uncertainty estimators with two main objectives: 1) achieve precise estimations of the probability of preference between two images in a pair, ensuring reliable results for the pairs that do not require human evaluation; 2) guide the pairwise sampling procedure to select the pairs that need human evaluation, thereby reducing the duration of a pairwise comparison subjective test.

\par The objective of this section is to introduce the proposed approach for uncertainty estimation. In this context, uncertainty estimation involves quantifying the uncertainty associated with the predictions made by a deep learning model. There are different ways to perform uncertainty estimation, which will be detailed in the following sections.

\subsection{Data Uncertainty}\label{Data_UN}
Aleatoric uncertainty, also referred to as data uncertainty, is an intrinsic characteristic of the input data distribution that denotes the inherent noise or variability, and it can be directly inferred from the data \cite{uncertaintySurvey}. This type of uncertainty is irreducible unless the data collection process is repeated. In the context of pairwise comparison assessment, the only method to mitigate data uncertainty (specifically, the variability in human decisions) is to directly request human observers to label the corresponding pair.

\par As in previous works \cite{UNIQUE}\cite{cao2024image}, it is assumed that the true perceptual quality $Q$ of an image follows a Gaussian distribution \cite{TM} with parameters $\mu$ and $\sigma$, which represent the mean and standard deviation of the quality, respectively. In the context of a pairwise comparison test, it is also assumed that the quality of the two images $A$ and $B$ of a pair, $Q(A)\sim \mathcal{N}(\mu_{A}, \sigma_{A})$ and $Q(B)\sim \mathcal{N}(\mu_{B}, \sigma_{B})$ are independent of each other and thus the quality difference $Q(A)-Q(B)$ distribution is also Gaussian as defined in \eqref{equ-diff_distribution}. 

\begin{equation}
\begin{split}
\label{equ-diff_distribution}
Q(A)-Q(B) \sim \mathcal{N}(\mu_{AB}, \sigma_{AB})
\quad\text{s.t.}\quad \\ 
\mu_{AB} = \mu_A - \mu_B,  \quad\sigma_{AB}^2 = \sigma_A^2 + \sigma_B^2
\end{split}
\end{equation}

The selection of image $A$ over image $B$ can be obtained by randomly sampling the quality difference $Q(A)-Q(B)$ distribution, and if the value obtained is greater than zero, image $A$ is selected over image $B$. This means that the positive area under the $Q(A)-Q(B)$ distribution indicates the probability of preferring $A$ over $B$, i.e. image $A$ has higher quality than image $B$. This could be quantified using the Gaussian cumulative distribution function (CDF) $\Phi()$ of the quality difference distribution, $Q(A)-Q(B)$ \cite{TM} as defined in \eqref{equ-preference}.
\begin{equation}
\label{equ-preference}
\hat{Pr}(A \succ B) = \Phi (\frac{ \mu_A - \mu_B}{\sqrt{\sigma_A^2 + \sigma_B^2}}) 
\end{equation}

where the $\succ$ symbol in \eqref{equ-preference} represent the preference and $\Phi(z)$ is CDF. 
\par The data uncertainty, which only depends on the input images, corresponds to how spread is the quality difference around the mean $\mu_{AB}$ and thus can be represented by the variance ($\sigma_{AB}^2$) of this distribution \cite{PC_analyze}, i.e. a high variance ($\sigma_{AB}^2$) represents a large dispersion and thus an increased level of uncertainty and lower variance indicates a distribution more centered around the mean, suggesting less uncertainty. 

\par
\begin{figure}[htbp]
\centerline{
  \begin{tabular}{@{}c@{}}
    {\includegraphics[scale=0.3]{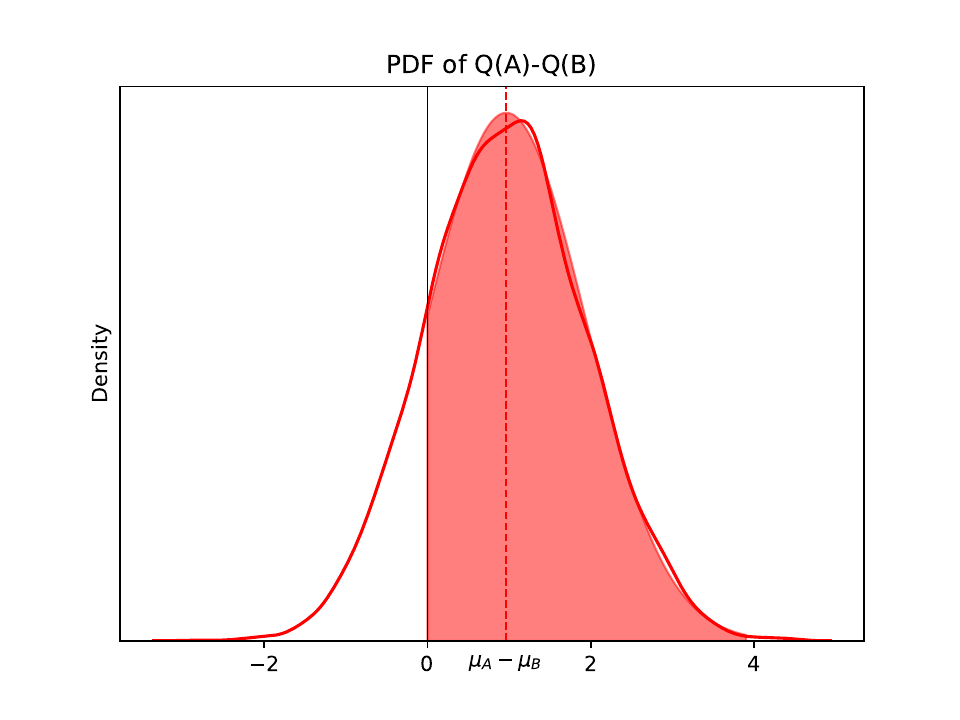}} \\ 
  \end{tabular}
}
\caption{Quality difference, $Q(A)-Q(B)$, distributions. The shaded area under the probability density function (PDF) curve of $Q(A)-Q(B)$ represents the probability of preference $\hat{Pr}(A \succ B)$.}
\label{fig1}
\end{figure}

 Figure \ref{fig1} provides a visual representation for quality difference distribution, where image $A$ exhibits higher quality than image $B$. Notably, the positive area under the quality difference distribution $Q(A)-Q(B)$ exceeds the negative area, indicating higher preference for selecting $A$ over $B$.
\vspace{-5pt}

\subsection{Model Uncertainty}\label{Model_UN}
\par Epistemic uncertainty, also known as model uncertainty, is an intrinsic characteristic of the model itself and arises from a lack of knowledge or data \cite{uncertaintySurvey}. This type of uncertainty can be mitigated by incorporating additional data or enhancing the model architecture and training process. Model uncertainty can influence the selection of pairs for human evaluation. When the model exhibits low confidence in its prediction of the preference between two images in a pair (denoted as $\hat{Pr}$ in \eqref{equ-preference}), model uncertainty is high. In such instances, the correct preference must be determined by human subjects, and the model’s prediction should be disregarded.

\par Monte Carlo dropout \cite{pmlr-v48-gal16}, \cite{Survey2}, referred to as MC-dropout, serves as an effective approximation of probabilistic Bayesian models within deep Gaussian processes. This technique involves applying dropout during training to mitigate overfitting by randomly deactivating some neurons in each training iteration. Consequently, a different subset of the network architecture is used each time. With MC-dropout, multiple passes are also executed during inference where neurons are disabled in each pass based on the outcome of a Bernoulli random variable (0 or 1) with a probability of $p$ \eqref{equ-bernoulli}. As a result, through $n$ forward passes of the same input, varying subsets of network architectures produce distinct outputs. Consider a dropout layer present in both training and test that is applied to the previous layer which has weights $W_i$, thus obtaining a new set of weights $D_i$ (after dropout) according to \eqref{equ-bernoulli}. 
\begin{equation}
\begin{gathered}
\label{equ-bernoulli}
    D_i = W_i \cdot diag([z_{ij}]_{j=1}^{k_i}) \\
     z_{i,j} \sim Bernoulli (p_i) \text{ for } i=1,...,L, j=1,...,k_{i-1}
\end{gathered}
\end{equation}
The binary variable $z_{i,j}$, used as input to layer $i$, indicates if unit $j$ in layer $i-1$ from tensor $W_i$ being dropped. $p_i$ is the activation probability for layer $i$ and is usually called dropout ratio and can be learned or set manually.
\par By repeatedly applying the same input, the model outputs can be interpreted as samples from a Gaussian distribution, providing an important characterization of the prediction’s variability. In fact, quantifying this distribution will allow uncertain inputs to be treated differently. The predictive Gaussian distribution $P_{m}$ is defined by parameters $\mu_{m}$ and $\sigma_{m}$, which are computed as $\mu_{m} = \frac{1}{n} \sum_{i=1}^{n}(\hat{Pr}(A \succ B))$ and $\sigma^2_{m} = Var(\hat{Pr}(A\succ B))$, respectively. In the context of pairwise sampling, the model uncertainty is therefore characterized by distribution $P_{m}$ serving two purposes: 
\begin{enumerate}[leftmargin=*]
    \item Enhance the robustness of predictions by averaging multiple stochastic forward passes. Thus, $\mu_{m}$ will be used as an improved estimation of the probability of preference.
    \item Quantification of the lack of confidence in the predicted preference between the two images of a pair. Higher values of $\sigma^2_{m}$ indicate greater dispersion around the mean, representing less reliable probability of preference estimation by the model. Thus, $\sigma^2_{m}$ will be used as a measure of reliability of the prediction.
\end{enumerate}

\section{Proposed Pairwise Sampling Procedure} \label{seq-pairwise-sampling-procedure}
This section outlines the primary contribution of the paper: the comprehensive learning-based pairwise sampling procedure that leverages uncertainty models to achieve precise preference estimations and identify the pairs that require subjective evaluation. To accomplish these goals, it is essential to detail the novel pairwise sampling framework (Section \ref{sampling-section}), the designed deep neural network model (Section \ref{net_arch}) and the pairwise selection algorithm (Section \ref{samp_proc}).

\subsection{Pairwise Sampling Framework}\label{sampling-section}

The proposed framework for pairwise sampling is illustrated in Fig. \ref{Fig_framework}, and summarized in the following steps:
\begin{figure}[htbp]
 \centering
 \includegraphics[scale=0.45]{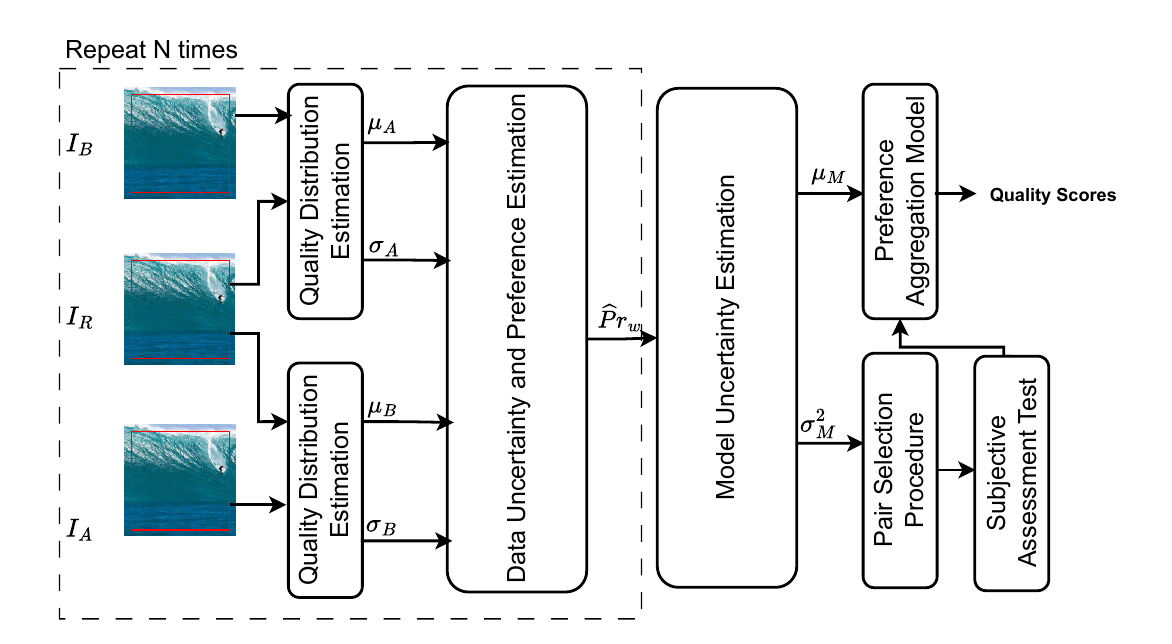}
 \caption{Overview of the pairwise sampling procedure.}
 \label{Fig_framework}
\end{figure}

\begin{enumerate}[leftmargin=*]
    \item \textbf{Quality distribution estimation}: first, the underlying quality distributions of both distorted images of a pair are calculated with a deep learning model presented in Section \ref{net_arch}. More precisely, two differentiable learned functions $f_{\mu}()$ and $f_{\sigma}()$, calculate the mean quality score and associated standard deviation using both the reference and each distorted image of the pairs. This step receives as input a pair of distorted images, denoted as $A$ and $B$ and the corresponding reference images $Ref$. The output is the quality distribution parameters for each of the distorted images of the pair. i.e. $\mu_A$, $\sigma_A$, $\mu_B$ and $\sigma_B$. 

    \item \textbf{Data uncertainty estimation}: the data uncertainty associated to the input images of the pair is then calculated according to \eqref{equ-diff_distribution}. This step receives as input all parameters calculated in step 1, and outputs the data uncertainty associated to the pair represented by $\sigma_{AB}^2$. In Section \ref{Data_UN}, data uncertainty was described in detail.

    \item \textbf{Preference estimation}: the probability of preferring image $A$ over image $B$ in a pair denoted as $\hat{Pr}(A \succ B)$ is calculated with \eqref{equ-preference}. This step receives as input $\mu_A$ and $\mu_B$ calculated in step 1 and the data uncertainty estimation of the previous step, i.e. $\sigma_{AB}^2$. The probability of preference $\hat{Pr}(A \succ B)$ is the output of this step. In Section \ref{Data_UN}, this step was described in detail.

    \item \textbf{Model uncertainty estimation}: the model uncertainty associated to the deep learning model is characterized by the Gaussian distribution $P_{m}$. This step receives as input a pair of images, repeatedly applies the same pair to the model (MC-dropout estimate) and computes the probability of preference (step 1 to 3). By running $n$ forward passes of a network with dropout in both functions $f_{\mu}()$ and $f_{\sigma}()$, $n$ predictions of $\hat{Pr}(A \succ B)$ are obtained since multiple estimates of $\mu_A$, $\sigma_A$, $\mu_B$ and $\sigma_B$ are provided by the neural network. This multiple passes allow to compute distribution $P_{m}$ and thus characterise the model uncertainty. The $\mu_m$ and $\sigma^2_m$ parameters of $P_{m}$ are used as a more reliable estimation of $\hat{Pr}(A \succ B)$ and as the dispersion (and thus confidence) in the model predictions, respectively. In Section \ref{Model_UN} model uncertainty was described in detail.

    \item \textbf{Pair selection procedure}: this procedure identifies the pairs for which the model is unreliable and should thus be evaluated by humans. The above steps are repeated for all possible pairs to calculate the data and model uncertainty associated with each one. All pairs are then sorted based on some specific criteria in descending order, and the top $X\%$ of the pairs are selected according to available budget. Three criteria are considered: two of those are rather straightforward, which is to sort according to $\sigma_{AB}^2$ or $\sigma^2_m$, i.e. from higher to lower data or model uncertainty, respectively. However, these two criteria do not consider the impact on the final scores of choosing a pair for subjective assessment. Therefore, a new criteria for pair selection based on expected information change (EIC) is proposed (refer to Section \ref{samp_proc} for details). In this step, the budget directly defines the length of the subjective test. 
     
    \item \textbf{Subjective assessment test}: finally, the pairs selected in the previous step undergo a subjective assessment test following a pairwise comparison methodology. Each paired comparison is evaluated by several subjects and thus the relative preference between two images can be obtained, i.e. probability of preference $Pr(A \succ B)$. 
    
    \item \textbf{Preference aggregation}: A pairwise comparison (PC) matrix is created and populated with $\mu_m$ of step 4 ($\hat{Pr}(A \succ B)$ estimate) and the $Pr(A \succ B)$ of the previous step (subjective score). Finally, the PC matrix is converted to scores using the Bradley-Terry preference aggregation model. The BT model parameters are estimated using maximum likelihood estimation and represent the preference strength of each image relative to the others. Images that are consistently preferred over others will receive higher scores. The output of this step is a quality score per image, which quantifies the relative attractiveness or preference for all distorted images.
\end{enumerate}

\begin{figure*}[htbp]
  \centering
  \includegraphics[scale=0.5]{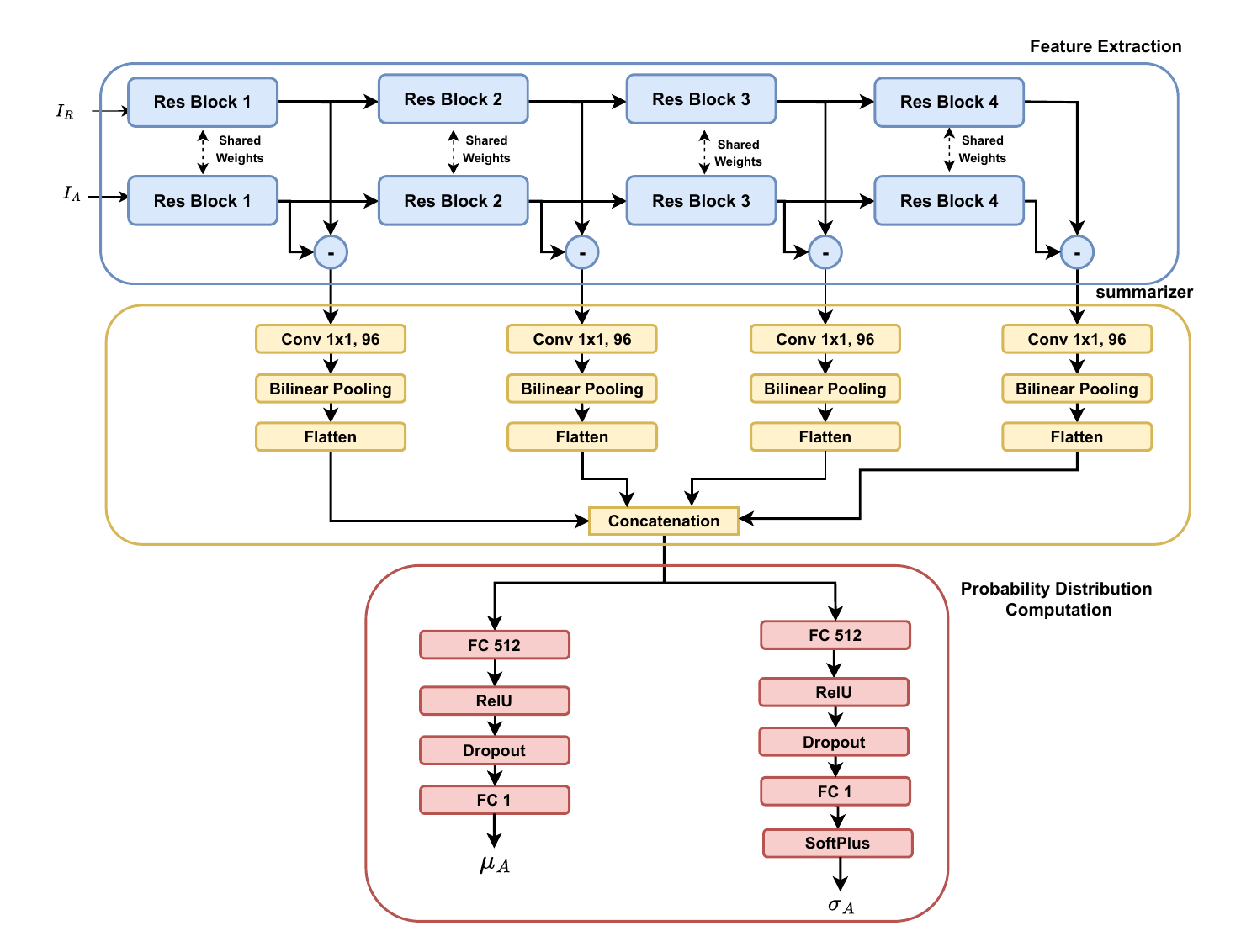}
  \caption{Proposed architecture of the deep learning framework ($f_{\mu}()$ and $f_{\sigma}()$).}
  \label{Fig_architecture}
  \vspace{-5pt}
\end{figure*}

\subsection{Network Architecture}\label{net_arch}
The function $f_{\mu}()$ and $f_{\sigma}()$ estimate the underlying quality distribution of each image of the pair using a deep network model. The proposed network architecture is shown in Fig. \ref{Fig_architecture} for image $I_A$, leveraging ResNet34 \cite{he2016deep} as the backbone for feature extractor. ResNet34 was selected as the backbone of this framework due to its balance between model complexity and performance. Its ability to extract robust, hierarchical features makes it effective at detecting subtle image distortions, which is crucial for quality assessment tasks. Additionally, ResNet34 is deep enough to capture complex patterns without being computationally prohibitive, offering both efficiency and scalability. The input is a distorted image (which can be $I_A$ or $I_B$) and the corresponding reference, $I_R$, and the output corresponds to the quality probability distribution parameters, $\mu$ and $\sigma$, of the distorted image relative to the reference. 

\par The weights of the feature extraction network for the distorted image are shared with the feature extraction network for the reference image. Residual maps are computed as the distance between the distorted image features and the reference image features after each residual block which facilitate the flow of gradient in back-propagation and leverage both high-level and low-level features (different scales of the input image) for the probability distribution computation. Each residual activation map then goes through a $1\times1$ convolution to enhance cross-channel interaction and obtain a smaller number of channels, which is important to reduce the number of parameters and computational cost of the fully connected layers at the end of the model. Note that features are obtained in different levels of the feature extractor network backbone (and thus different resolutions) but the number of output channels is the same for all $1\times1$ convolutions. After, bilinear pooling \cite{bilinearPooling} is applied to capture second-order statistics and pairwise feature interactions according to \eqref{equ-bilinear}, resulting in more detailed and discriminative set of features, which is essential for accurately estimate the probability distribution parameters for each image. Before bilinear pooling, spatial, $s$, and then channel, $c$, dimensions are flattened to obtain a vector $z\in \mathbb{R}^{s\times c}$. The bilinear pooling is defined in \eqref{equ-bilinear}.
\begin{equation}
\label{equ-bilinear}
    \hat Z = z^T z
\end{equation}
After flattening and concatenation, the features are passed through function $f_{\mu}()$ which is represented by a fully connected layer with 512 units and the ReLU non-linear activation. Following the model uncertainty approach described in \ref{Model_UN}, a dropout layer was introduced which is applied both during training and inference. In this case, dropout randomly ''drops out'' (sets to zero) a fraction of the neurons in the previous layer. This means that these neurons are temporarily removed from the network, along with all their incoming and outgoing connections. The dropped neurons are selected at random with a probability $p$, known as the dropout rate. After, another fully connected layer with a single unit is used to generate the mean, $\mu$, of the Gaussian distribution parameter. 

\par Regarding, function $f_{\sigma}()$, the concatenated feature maps are passed through a fully connected layer with 512 outputs and a ReLU activation, followed by dropout as in function $f_{\mu}()$. Since both functions $f()$ have dropout layers, the probability of preference is estimated in multiple passes using different values of $\mu$ and $\sigma$; this enables the computation of model uncertainty. Then, another fully connected layer with a single output and a SoftPlus activation is used to constrain the values of $\sigma$ to be positive while avoiding the problem of having zero gradients for negative inputs (also being more smooth than ReLU around zero). Moreover, in Section II of the supplementary material, all details of the deep learning network architecture are presented.

\subsection{Pair Selection Procedure}\label{samp_proc}
This section introduces a novel procedure to select pairs aimed at enhancing the accuracy of scores obtained after preference aggregation. An expected information change criterion is proposed, measuring the impact on the final subjective scores when a pair is chosen for subjective evaluation. The sampling algorithm relies on the computation of the posterior distribution over the scores obtained from selecting a pair for subjective assessment. This criterion is based on the typical utility function of \cite{ASAP}\cite{Hybrid-MST}, where the expected information gain (EIG) is computed by considering the two possible choices when a subject evaluate a pair. The posterior distribution can help to select the next comparison by accounting for possible deviations in the estimated probability of preference according to the model uncertainty. 
\par Initially, a pairwise comparison matrix $M$ is created and populated with the mean predicted preferences $\mu_m$ obtained from the model uncertainty step. This is the initial state and corresponds to the probability of preference estimated from the score distributions obtained by the deep network model presented in Section \ref{net_arch}. Next, the prior distribution of quality scores $p(S|M)$ are derived from $M$ using the Bradley-Terry algorithm. To account for potential discrepancies, the posterior distribution $P_{ij}^{\pm}(S|M,A)$ is computed for each pair $(i,j)$ by updating the predicted preferences of $M$ with $A_{ij}$ which is calculated according to \eqref{equ-delta}.
\begin{equation}
    \label{equ-delta}
    A_{ij} = \text{Clip}(M(i,j) \pm \max(\delta, \sigma_m^2(i,j)),0,1)
\end{equation}
Since predicting the outcome of selecting a pair is not possible, two possible cases are considered for pair $(i,j)$: i) increasing or ii) decreasing $M(i,j)$ by some amount. This amount should reflect the model’s uncertainty about its probability of preference estimation. In this work, it was found that the update provided by \eqref{equ-delta} works best, i.e. adjusting the preference by $\pm \delta$ or the model uncertainty $\sigma_m^2(i,j)$, which is first normalized using min-max normalization. When adjusting the preferences of $M$, the results are clipped with $\text{clip}(x, a, b) = \max(a, \min(x, b))$ to ensure the new values remain valid probabilities between zero and one, as specified in \eqref{equ-delta}. Next, the Expected Information Change (EIC) for each pair is calculated according to \eqref{equ-kld}. In this context, the EIC corresponds to the Kullback-Leibler (KL) divergence $D_{KL}$  between the prior and posterior scores, which are modeled as multivariate normal distributions.

\begin{equation}
\begin{split}
    \label{equ-kld}
    EIC(i, j) = D_{KL}(p(S|M)\:||\: P^{+}_{i,j}(S|M,A)) + \\ 
    D_{KL}(p(S|M)\:||\: P^{-}_{i,j}(S|M,A))
\end{split}
\end{equation}
This process is repeated for every pair of images $(i,j)$ separately, and pairs expected to yield the maximum EIC are selected, i.e. pairs are sorted from the maximum to the minimum EIC. This approach ensures that the most impactful pairs are selected, improving the accuracy and robustness of the final quality score predictions.


\section{Performance Evaluation} \label{performance-section}
In this section, the proposed solution is thoroughly evaluated. First, the training procedure of the proposed deep learning model is presented, followed by the dataset used and the test conditions (for inference). Then, the experimental results are presented, namely for the evaluation of the deep learning model, pair selection and comparison with state of the art. From now on, the proposed \textbf{L}earning-\textbf{B}ased \textbf{P}airwise \textbf{S}ampling is referred to as LBPS-Data for data-based sampling, LBPS-Model for model-based sampling, and LBPS-EIC for EIC-based sampling.

\subsection{Training Procedure}
\par The objective of the training procedure is to find network parameters $\theta$, that given the input images $A$, $B$ and $Ref$, can accurately characterize the quality of both images of a pair. The input images are first center cropped to $224\times224$ as typical when the ResNet network architecture is used. The network parameters are initialized with the weights trained on ImageNet \cite{imageNet}. The training hyper parameters were carefully selected to optimize performance, namely, 8 pairs of images, $A$, $B$ and $Ref$, were chosen randomly for each batch, and the learning rate was initialized to $1E-5$. A scheduler reduces the learning rate by a factor of 0.1 after each 10 epochs to improve convergence and preventing over-fitting. The dropout ratio used for training but also for inference is 0.2.

The fidelity loss function in this paper is particularly effective for handling probability distributions, as it quantifies the distance between probabilities and is bounded between 0 and 1, making it stable, efficient, and interpretable. Inspired by quantum physics, this loss function outperforms existing methods in experimental tests on a large-scale dataset within a probabilistic ranking framework \cite{fidelity_loss}. The fidelity loss function is defined in \eqref{equ-loss_func} and measures the similarity between two probability functions: namely $\Pr (A \succ B)$, the probability of image $A$ being preferred over $B$ from the ground truth dataset, and the corresponding probability $\hat Pr(A \succ B)$ estimated using \eqref{equ-preference}, the distribution parameters estimated by the proposed neural network model. 

\begin{equation}
\begin{split}
\label{equ-loss_func}
    Loss = 1-\sqrt{Pr(A\succ B)\hat Pr(A\succ B)} \\
    - \sqrt{(1-Pr(A\succ B)(1-\hat Pr(A\succ B))}
\end{split}
\end{equation}

\subsection{Training and Test Datasets}
\par The proposed architecture was trained using PieAPP dataset \cite{Prashnani_2018_CVPR} which consists of three disjoint sets for train, validation and test all with a resolution of $224\times224$. The training and validation set consist of 140, and 20 reference contents respectively with 123 degraded images associated with each reference. Specifically, the PC subjective test was designed for two kinds of pairs: inter-pairs and intra-pairs, which represent different distortion scenarios for each pair of images. In an inter-type comparison, A and B have different types of distortions. However, in intra-type comparison, A and B have the same type of distortions. The training set comprises 77k image pairs, and the validation set includes 9k pairs. Additionally, the test set consists of 40 reference images, each subjected to 15 different degradation types. The degradation types present in the test set differ entirely from those in the training and validation sets. This dataset configuration ensures diverse training examples, and assesses the model’s ability to generalize to unseen degradation types. Each example in the aforementioned sets is labeled with the probability of preference, $Pr(A\succ B)$ obtained through a crowd-sourcing based PC subjective test in which subjects were asked to select the image that is more similar to the reference image. 

\par Additionally, the proposed LBPS solutions were evaluated using the Pairwise Comparison Image Quality Assessment (PC-IQA) dataset \cite{Xu2012}. This dataset was collected through a crowd-sourced pairwise subjective test methodology and includes 15 reference images, each with 15 corresponding distorted versions, all at a resolution of $480\times720$. As a result, the dataset comprises 1,800 image pairs, with 120 pairs per reference image. While the dataset is comprehensive, comparing all possible pairs, it is imbalanced because each pair has been assessed a different number of times.

\subsection{Test Procedure}\label{sec:test_proc}

To assess the proposed framework, a simple algorithm is employed that utilizes several pairwise comparison matrices (PCMs) $\bm{m}$. Each PCM captures the preferences of a set of images linked to a specific reference (or content) \cite{ISM2022}. This set of images comprises the original image and its degraded versions. Thus, the test procedure begins with the creation of the $\bm{m}_i$ PCMs which are populated with the preferences of the ground-truth for reference $i$. After, $\bm{\hat{m}}_i$ PCMs are created and populated with estimated preferences for a particular method. For instance, for LBPS-Data, these preferences are $\hat{Pr}(A \succ B)$ and for LBPS-Model and LBPS-EIC, the preferences are $\mu_{m}$. Moreover, some preferences in $\bm{\hat{m}}_i$ may have been obtained from ground truth scores if the pair was selected for human evaluation. After, the BT preference aggregation model is applied to each $\bm{m}_i$ to infer scores $\bm{s}_i$ and to each $\bm{\hat{m}}_i$ to obtain $\bm{\hat{s}}_i$. The scores $\bm{s}_i$ and $\bm{\hat{s}}_i$ for all the references $i$ are concatenated and some metric is applied to compute the correlation between $\bm{s}$ and $\bm{\hat{s}}$. This allows for the assessment of the quality scores produced for the entire dataset and not just for one reference.

\par In this work, three correlation metrics are employed: Pearson Linear Correlation Coefficient (PLCC), Spearman Rank-Order Correlation Coefficient (SROCC), and Root Mean Square Error (RMSE). PLCC is utilized to measure the linear correlation between the predicted and the ground truth quality scores, indicating how closely the predicted scores align with the ground truth scores. SROCC assesses the strength and direction of the monotonic relationship between the predicted and ground truth quality scores, providing insight into the consistency of the ranking order produced by the model relative to the ground truth. RMSE, meanwhile, quantifies the average magnitude of the error between the predicted and actual quality scores providing a direct measure of prediction accuracy. 

\subsection{Anchors}\label{sec:anchors}

The proposed LBPS-EIC solution was evaluated against several state-of-the-art methods to assess its performance and robustness, including our previous work, PS-PC. Random sampling is the most straightforward solution and serves as baseline. In such case, the initial preference is set to 0.5, indicating no preference, and pairs are selected randomly for comparison. The random sampling process is repeated 25 times, and the average performance is reported. Moreover, in this evaluation, PS-PC uses four classifier/predictor models, which results in fixed number of selected pairs were used. Other active sampling approaches were included as anchors, particularly those more efficient, such as ASAP \cite{ASAP}, Crowd-BT \cite{Crowd-BT}, Hybrid-MST \cite{Hybrid-MST} and HR-Active \cite{HrActive}, all with the highest performing configuration. Active sampling approaches function differently in that they do not select pairs directly. Instead, they choose a trial (a comparison) or a set of trials in each run of the algorithm. Consequently, it requires raw judgments from subjects for each pair, which are not available in some datasets. For instance, the PieAPP dataset only provides the probability of preference between all possible pairs of images, requiring the creation of binary judgments to be used in the evaluation of active sampling methods.

\par Therefore, for the PieAPP dataset, the observed judgments were simulated based on the probability of preference available for each pair of images. A Bernoulli distribution was used, assuming that individual trials are independent, where the success probability, $p$, represents the probability of preferring one image over the other. Sampling the Bernoulli distribution involves generating a random variable (using a uniform distribution) and choosing image $A$ over image $B$ if the generated random number is less than the ground truth preference. Due to inherent randomness in generating binary judgments, 25 iterations of the active sampling approaches were conducted, and the average is reported. Moreover, the PC matrix was initialized with a probability of preference 0.5 for the anchors. 

\par Additionally, the proposed LBPS-EIC solution was evaluated on PC-IQA, which binary judgements are available in the dataset, to further assess the performance. It is important to note that anchors randomly select a judgment from all the available judgments. Since these decisions are obtained randomly, an average of 25 iterations is used.

\subsection{Experimental Results}
This Section presents the several experiments made to evaluate the main contributions of this work, namely the performance of the deep learning model (Section \ref{sec-DNN-performance}), the proposed pair selection method (Section \ref{sec-sampling_eval}), the evaluation of the proposed solution with the highest performance, against the anchors previously defined (Section \ref{sec-sota}) and confidence analysis of the proposed solution (Section \ref{variance-analysis}). A qualitative analysis with a focus on high uncertainty cases is shown in Section I of the supplementary material. In all the following experiments, 200 iterations was used for MC-Dropout, and $\delta$ is equal to 0.3 in the LBPS-EIC criteria.

\subsubsection{Deep Learning Model Evaluation}
\label{sec-DNN-performance}
In this Section, the probability of preference $\hat{Pr}(A \succ B)$ computed for all pairs from the quality distributions estimated by the deep neural network model is analyzed in two different experiments. 

\par
\begin{figure}[htbp]
\centerline{
\begin{tabular}{@{}c@{}}
\vspace{-4pt}
    {\includegraphics[scale=0.31]{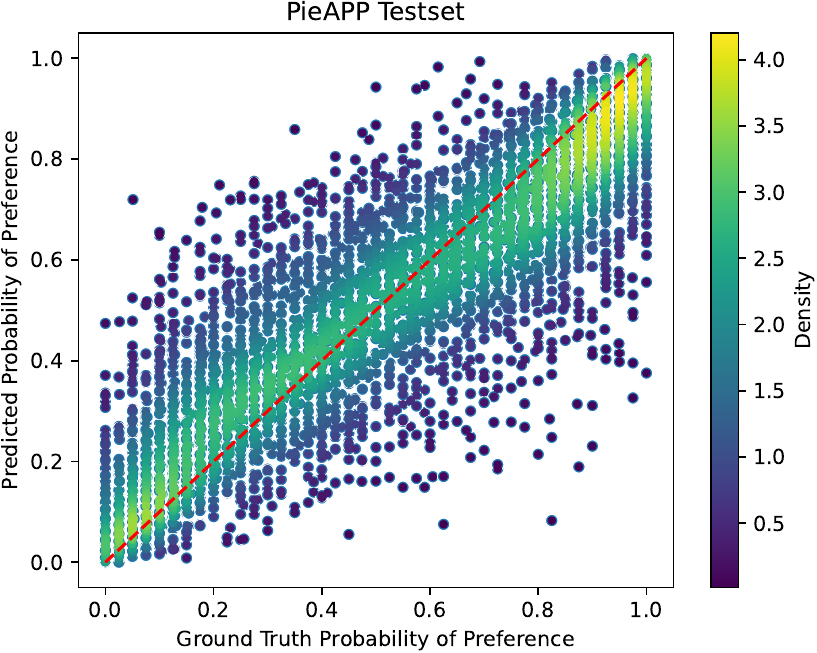}} \\
  \end{tabular}
\begin{tabular}{@{}c@{}}

    {\includegraphics[scale=0.31]{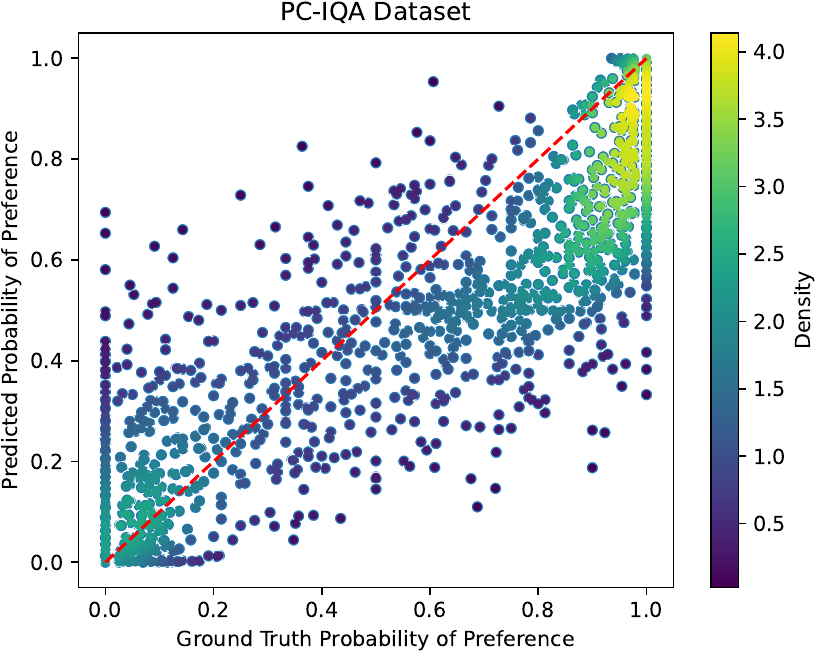}} \\
  \end{tabular}}
\caption{Predicted preference versus ground truth preference using a scatter plot combined with a heatmap. The heatmap overlays the scatter plot to illustrate the density of data points.}  
\label{Fig_predicted_pref}
\end{figure}

 First, Fig. \ref{Fig_predicted_pref} illustrates the relationship between the predicted $\hat{Pr}(A \succ B)$ and ground truth preferences $Pr(A \succ B)$ for all pairs of the PieAPP test set, and PC-IQA dataset, providing a clear and comprehensive visual summary of the model's performance. As shown, the data points are clustered near the $y=x$ line which indicates the model’s strong performance in estimating the probability of preference.

\vspace{-6pt}

\begin{figure}[htbp]
\centerline{
\begin{tabular}{@{}c@{}}
    {\includegraphics[scale=0.31]{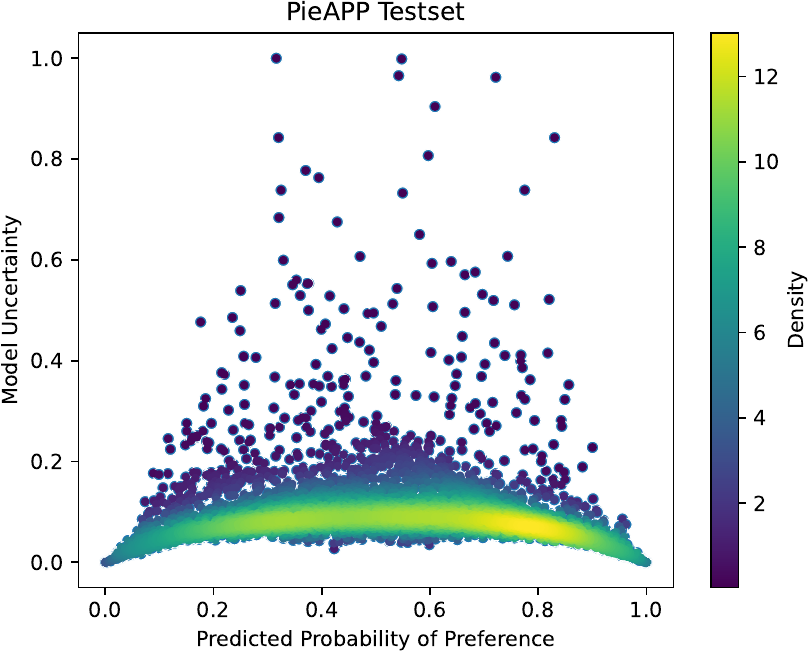}} \\
  \end{tabular}
\begin{tabular}{@{}c@{}}

    {\includegraphics[scale=0.31]{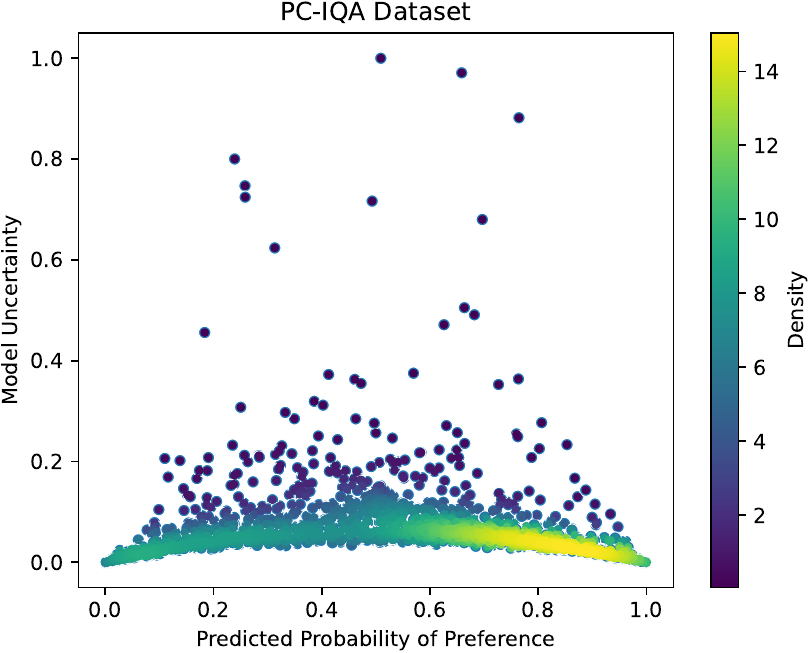}} \\
  \end{tabular}}
\caption{Model uncertainty versus predicted preferences.}  
\label{Fig_model_vs_pref}
\end{figure}

Fig. \ref{Fig_model_vs_pref} illustrates the relationship between the predicted preferences, denoted as $\hat{Pr}(A \succ B)$, and the estimated model uncertainty, $\sigma^2_m$. As shown, model uncertainty is higher when the images of a pair have similar quality levels, which means probability of preference around 0.5. Moreover, when one of the images has significantly higher or lower quality than the other (probability of preference close to 1 or 0), model uncertainty decreases as expected. 

\subsubsection{Pair Selection Evaluation} \label{sec-sampling_eval}

\begin{figure*}
    \centering
    \includegraphics[scale=0.4]{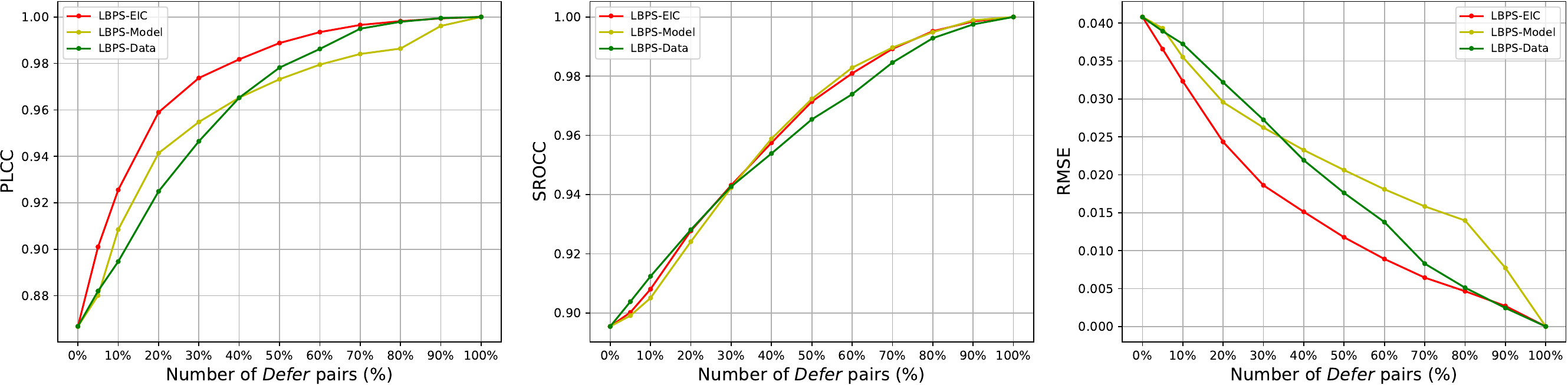}
    \caption{Evaluation of the pairwise selection method}
    \label{fig:Un}
\end{figure*}

In this Section, the LBPS-Data, LBPS-Model, LBPS-EIC pair selection methods are evaluated. Fig. \ref{fig:Un} presents the experimental results for all methods using the three performance metrics PLCC, RMSE and SROCC following the test procedure of Section \ref{sec:anchors}. As expected, the overall performance improves with an increasing budget of selected pairs (denoted as $X\%$) for subjective assessment. When no pairs are selected (0\%), the performance reflects the model’s predictive capability, resulting in identical performance across all criteria.

\par However, when the budget increases, LBPS-EIC, which leverages both data and model uncertainties, achieves the highest performance. Specifically, selecting 10\% of pairs yields a PLCC correlation above 0.92, which rises to nearly 0.99 at 50\% selection. LBPS-Model demonstrates the second-best performance when fewer than 40\% of pairs are selected, but its performance lowers compared to LBPS-Data as the number of deferred pairs increases. This means that the uncertainty associated to the data (and not in the sampling framework) is the best driving force to select pairs, which is expected when many pairs are selected and the possible errors made by the neural network model in its estimation were identified and corrected. A similar behavior could also be observed for the RMSE metric, where LBPS-EIC clearly outperforms other approaches.

\par Regarding the SROCC metric, overall performance also improves with an increasing number of selected pairs for subjective assessment. However, all three criteria exhibit similar performance, indicating minimal change in the rank order of predictions or minor misordering. This suggests that while all uncertain estimation models enhance the fidelity of the scores, they still maintain the ranking of the stimuli. 

\par In conclusion, these findings underscore the effectiveness of LBPS-EIC by incorporating both data and model uncertainties, particularly in scenarios involving a low to medium budget of selected pairs for subjective assessment.

\subsubsection{State-of-the-art Comparison}\label{sec-sota}

In this section, the best-performing method, LBPS-EIC, is evaluated in comparison to the anchors, following the test procedure outlined in Section \ref{sec:test_proc}. To facilitate this evaluation, a fixed number of trials (budget) was selected to evaluate all methods: 2.5\%, 5\%, 7.5\%, 10\%, 15\%, 20\%, 30\%, 40\% and 50\%. Here, the budget defines the number of trials that can be conducted, expressed as a percentage. The maximum number of trials is set to $\frac{n(n-1)}{2} \times 15$, with 15 representing the number of subjects. For the anchors, the average over 25 runs is reported and the standard deviation over these runs is not significant and thus is not reported. The performance of the LBPS-EIC is compared with selected anchors using all test datasets. Experimental results for the PS-PC method were not included for the PC-IQA dataset since the training (used for training PS-PC) and test datasets are rather similar due to the use of the same degradation types.

\par As shown in Table \ref{tbl-SOTA-PieAPP} and Table \ref{tbl-SOTA-PC-IQA}, LBPS-EIC demonstrates notable performance improvements compared to random sampling across both datasets. For a budget of 50\% trials, LBPS-EIC achieves a PLCC correlation of 0.99/0.98 for the PieApp/PC-IQA datasets, compared to 0.83/0.66 for random sampling. This indicates that at 50\% of the trials, the scores obtained after pair selection with LBPS-EIC are already very similar to the ground truth scores. Furthermore, as shown in Table \ref{tbl-SOTA-PieAPP} and Table \ref{tbl-SOTA-PC-IQA}, LBPS-EIC consistently exhibits the highest PLCC correlation across all trial budgets compared to other methods for both datasets. For the PieApp dataset, at 10\% of trials, LBPS-EIC achieves a PLCC correlation of 0.93, whereas the anchors yield lower correlations of 0.82, 0.79, 0.86, and 0.80 for ASAP, Crowd-BT, Hybrid-MST, and PS-PC, respectively. For the PC-IQA dataset a similar behaviour is observed, at $10\%$ of trials, LBPS-EIC achieves a PLCC correlation of 0.91, whereas the anchors have much lower correlations between 0.67 to 0.86. 

\par Regarding, the SROCC correlation metric, Hybrid-MST has the highest correlation, albeit only slightly superior compared to LBPS-EIC. The difference between SROCC values is less than 0.03 for the PieApp dataset and 0.04 for the PC-IQA and only occurs for budgets greater than 10\%. However, LBPS-EIC continues to show improvements in correlation performance as the trial budget increases and has almost the same performance when 50\% of the pairs are selected. Note also, that for a budget less than 10\%, LBPS-EIC achieves the best performance since its able to provide accurate estimation of the probability of preference and selects the most impactfull pairs. An important point to note is that active sampling methods that are used as anchor require running the sampling procedure after each trial (or a batch of trials), which is challenging in many scenarios (e.g., crowd-sourcing) where many subjects evaluate pairs in parallel or the computational resources to select the pairs are insufficient. For instance, if four subjects are doing the test in parallel the selection performance of Hybrid-MST batch mode will be lower since all the data is only obtained (and thus can be used to compute the next pairs) after the four subjects finish their evaluations. In the proposed solutions, this procedure is much more simplified since the selection of pairs is done \textit{a priori} of the subjective test.

\par With respect to the RMSE correlation metric, LBPS-EIC consistently outperforms the other methods across all trial budgets, further emphasizing its effectiveness compared to the anchor methods. Actually, the performance gains with RMSE are rather significant, which means that the underlying quality distributions of both distorted images of a pair calculated with the proposed deep learning model are accurate. These quality distributions are used to estimate the probability of preference which are then used by the Bradley-Terry preference aggregation model to obtain the scores used for performance evaluation.

\par The differences in results between the PieAPP and PC-IQA datasets, particularly at lower trial percentages, can primarily be attributed to variations in image resolution and the cropping method used during evaluation. The PieAPP images, with a resolution of $256\times256$, retain most of their visual information when cropped to $224\times224$. On the other hand, the higher-resolution images in the PC-IQA dataset ($480\times720$) experience a more significant loss of detail when cropped to $224\times224$, which may influence the assessment results. Although higher resolution provides more visual detail, cropping is essential for practical subjective pairwise testing, as it ensures images can be displayed side-by-side without distortion, adhering to screen resolution limitations.
\par In summary, while our method demonstrates comparable or even superior performance to existing approaches in terms of correlation metrics like PLCC, SROCC, RMSE or the number of pairs selected, its primary advantage lies in its offline operation. This capability enhances efficiency and scalability, making it particularly suited for large-scale subjective evaluations where minimizing runtime complexity is critical.

\subsubsection{Confidence Analysis}\label{variance-analysis}
\begin{figure}
    \centering
    \includegraphics[width=1\linewidth]{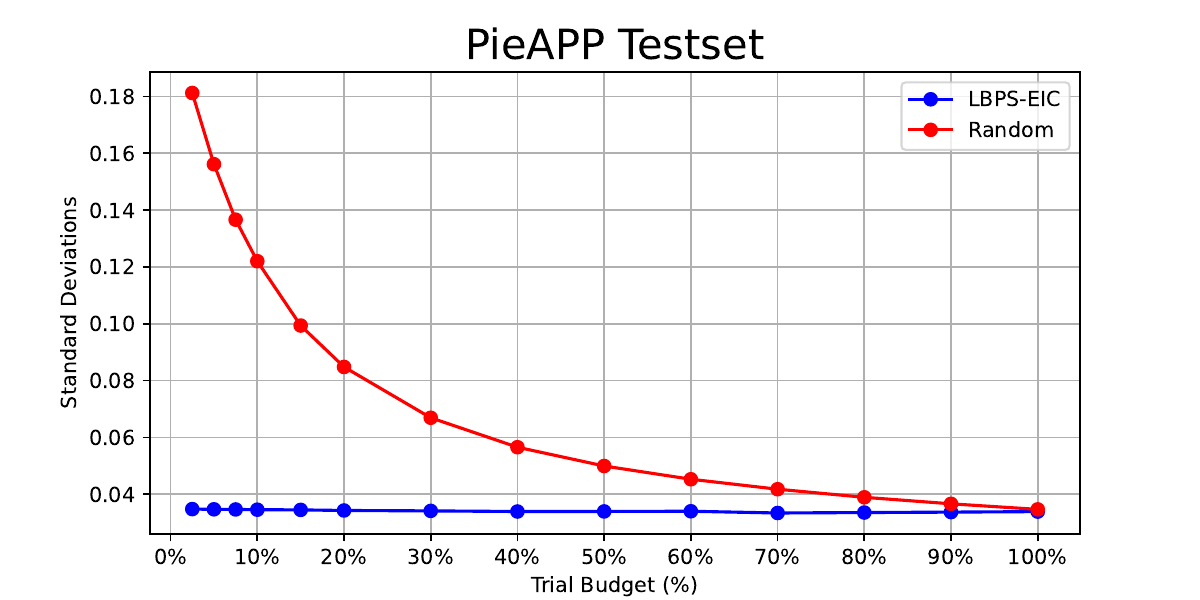}
    \caption{Performance evaluation reporting standard deviation over the entire test dataset.}
    \label{fig:varinace_pieAPP}
\end{figure}

This Section presents a confidence analysis of the scores obtained after pairwise sampling. This is complementary since correlation metrics between the ground-truth scores (usually obtained using a complete/exhaustive design) and the scores obtained with the sampling-based approach may not fully characterize the performance of the proposed solution. To assess confidence (or reliability), the average standard deviation of the inferred quality scores is calculated using the Bradley-Terry preference aggregation model, as outlined in Section \ref{sampling-section}. In this case, both the scores and standard deviations are obtained using maximum likelihood estimation. This standard deviation is averaged across the entire dataset for each trial budget. The proposed LBPS-EIC approach is compared to random selection, a sampling method that can also be performed before the subjective test starts. As shown in Fig. \ref{fig:varinace_pieAPP}, the average standard deviation for random selection begins at 0.18 and decreases as the number of deferred pairs increases, as expected. On the other hand, the standard deviation for the proposed approach remains small and consistent across all percentages, demonstrating its robustness and reliability. Moreover, the experimental results obtained highlight the importance of carefully selecting the best pairs in a pairwise comparison subjective evaluation. 

\begin{table*}[htb]
\scriptsize
\caption{LPBS-EIC performance evaluation compared to anchors for the PieAPP test set.}
\centering
\scalebox{1.5}{
\begin{tabular}{l|c|c|c|c|c|c|c|c|c|c}
\hline
\specialrule{0.2em}{\abovetopsep}{\belowbottomsep}

\multirow{2}{*}{Method} & \multirow{2}{*}{Metric} & \multicolumn{9}{c}{Trial budget (\%)} \\
\cline{3-11}
&    & $2.5\%$    & $5\%$   & $7.5\%$  &  $10\%$     & $15\%$ & $20\%$  & $30\%$  & $40\%$  &$50\%$       \\
\hline
\specialrule{0.2em}{\abovetopsep}{\belowbottomsep}

\multirow{3}{*}{LBPS-EIC}    &PLCC   &\textbf{0.89} &\textbf{0.90} &\textbf{0.92} &\textbf{0.93} & \textbf{0.94}  &\textbf{0.96}   &\textbf{0.97}   &\textbf{0.98}  &\textbf{0.99}        \\
                            &SROCC  &\textbf{0.89} &\textbf{0.90} &\textbf{0.90} &0.91 &0.92  &0.93  &0.94   &0.96  &0.97      \\ 
                            &RMSE &\textbf{0.038} &\textbf{0.036}  &\textbf{0.033} &\textbf{0.031} &\textbf{0.028} &\textbf{0.023} &\textbf{0.018} &\textbf{0.015} &\textbf{0.011}\\

\hline

\multirow{3}{*}{ASAP}        &PLCC    &0.55 &0.71 &0.79  &0.82 &0.87  &0.89   &0.92   &0.94  &0.95         \\
                               &SROCC &0.73 &0.85 &0.90  &0.92 &0.94 &0.96   &0.97   &0.98  &0.98         \\
                               &RMSE &0.069  &0.064  &0.060 &0.057 &0.052 &0.048  &0.042 &0.037  &0.034 \\
                               
\hline

\multirow{3}{*}{Crowd-BT}        &PLCC    &0.53 &0.67 &0.74  &0.79 &0.83  &0.86   &0.89   &0.90  &0.91          \\
                               &SROCC  &0.64 &0.78 &0.85  &0.89 &0.93 &0.94   &0.95   &0.96  &0.96        \\
                               &RMSE &0.069 &0.064 &0.058 &0.055 &0.053 &0.050 &0.048 &0.046 &0.045 \\

\hline

\multirow{3}{*}{Hybrid-MST}        &PLCC    &0.67 &0.78 &0.83  &0.86 &0.90   &0.92   &0.94   &0.95  &0.95        \\
                               &SROCC  &0.77 &0.87  &0.90 &\textbf{0.92} &\textbf{0.95}  &\textbf{0.96}   &\textbf{0.97}   &\textbf{0.98}  &\textbf{0.98}          \\
                               &RMSE &0.064 &0.058 &0.053 &0.050 &0.044 &0.040 &0.034 &0.030 &0.028 \\
                              
\hline

\multirow{3}{*}{HR-Active}        &PLCC    &0.65 &0.74 &0.78  &0.81 &0.84   &0.87   &0.89  &0.91  &0.92        \\

                               &SROCC  &0.76 &0.84  &0.87 &0.89 &0.92  &0.94   &0.95   &0.96  &0.97          \\
                               
                               &RMSE &0.069 &0.066 &0.064 &0.06 &0.060 &0.057 &0.053 &0.049 &0.046 \\
                              
\hline

\hline

\multirow{3}{*}{PS-PC}        &PLCC    &- &- &-  &0.80 &-  &0.84   &0.86   &0.89  &0.89         \\
                            &SROCC  &- &- &-  &0.44 &-  &0.49   &0.53   &0.60  &0.62           \\
                            &RMSE &- &- &-  &0.048 &-  &0.042   &0.040   &0.034  &0.035\\ 
\hline

\multirow{3}{*}{Random}        &PLCC    &0.30 &0.42 &0.49  &0.55 & 0.61 &0.66   &0.74   &0.78  &0.83          \\
                            &SROCC  &0.31 &0.48 &0.6  &0.67 &0.75  &0.81 &0.87   &0.91  &0.93         \\
                           &RMSE &0.075 &0.074 &0.073 &0.072 &0.070 &0.068 &0.064 &0.059 &0.053 \\
\hline

\specialrule{0.2em}{\abovetopsep}{\belowbottomsep}
\end{tabular}
}

\label{tbl-SOTA-PieAPP}
\end{table*}

\begin{table*}[hbt!]
\vspace{-10pt}
\scriptsize
\caption{LPBS-EIC performance evaluation compared to anchors for the PC-IQA test set.}
\centering
\scalebox{1.5}{
\begin{tabular}{l|c|c|c|c|c|c|c|c|c|c}
\hline
\specialrule{0.2em}{\abovetopsep}{\belowbottomsep}

\multirow{2}{*}{Method} & \multirow{2}{*}{Metric} & \multicolumn{9}{c}{Trial budget (\%)} \\
\cline{3-11}
&    & $2.5\%$ & $5\%$  & $7.5\%$ & $10\%$ & $15\%$ & $20\%$  & $30\%$  & $40\%$  &$50\%$      \\
\hline
\specialrule{0.2em}{\abovetopsep}{\belowbottomsep}

\multirow{2}{*}{LBPS-EIC}    &PLCC   &\textbf{0.83} &\textbf{0.87} &\textbf{0.89} &\textbf{0.91} &\textbf{0.93} &\textbf{0.95}   &\textbf{0.96}  &\textbf{0.98}  &\textbf{0.98}         \\
                            &SROCC  &\textbf{0.83} &\textbf{0.84} &0.85  &0.85 &0.87   &0.9   &0.92  &0.95 &0.96      \\
                            &RMSE &\textbf{0.088} &\textbf{0.083} &\textbf{0.081} &\textbf{0.074} &\textbf{0.060} &\textbf{0.053} &\textbf{0.042} &\textbf{0.031} &\textbf{0.025} \\

\hline

\multirow{2}{*}{ASAP}         &PLCC    &0.45 &0.60 &0.70  &0.74 &0.79  &0.83   &0.88  &0.92  &0.94       \\
                               &SROCC  &0.73 &0.82 &0.85  &0.86 &0.88  &0.89   &0.91  &0.92 &0.93      \\
                               &RMSE &0.133 &0.128 &0.0123 &0.120 &0.113 &0.106 &0.094 &0.084 &0.076 \\
                               
\hline

\multirow{2}{*}{Crowd-BT}        &PLCC    &0.44 &0.56 &0.62  &0.67 &0.73  &0.78   &0.83  &0.86  &0.88        \\
                               &SROCC  &0.62 &0.74 &0.80  &0.83 &0.85  &0.87   &0.89  &0.9 &0.90      \\
                               &RMSE &0.133 &0.128 &0.124 &0.122 &0.118 &0.115 &0.109 &0.105 &0.101 \\

\hline

\multirow{2}{*}{Hybrid-MST}        &PLCC    &0.66 &0.78 &0.83   &0.86 &0.91  &0.94   &\textbf{0.96}  &0.97    &0.99    \\
                               &SROCC  &0.79 &0.85 &\textbf{0.87}  &\textbf{0.89} &\textbf{0.91}   &\textbf{0.92}   &\textbf{0.94}  &\textbf{0.96} &\textbf{0.96}    \\
                               &RMSE &0.127 &0.119 &0.112 &0.105 &0.093 &0.082 &0.065 &0.052 &0.043 \\
                              
\hline

\multirow{3}{*}{HR-Active}        
                                &PLCC    &0.56 &0.65 &0.7  &0.73 &0.77   &0.80   &0.84  &0.87  &0.89        \\

                               &SROCC  &0.76 &0.79  &0.82 &0.83 &0.85  &0.86   &0.87   &0.88  &0.89          \\
                               
                               &RMSE &0.13 &0.131 &0.129 &0.127 &0.123 &0.12 &0.113 &0.107 &0.101 \\
                              
\hline

\hline


\multirow{2}{*}{Random}        &PLCC    &0.23 &0.32 &0.35 &0.40 &0.46 &0.50   &0.57   &0.63  &0.66        \\
                            &SROCC  &0.31 &0.44 &0.54 &0.58 &0.67 &0.71  &0.77   &0.81  &0.83           \\
                            &RMSE &0.136 &0.135 &0.134 &0.133 &0.131 &0.129 &0.125 &0.120 &0.114 \\

\hline

\specialrule{0.2em}{\abovetopsep}{\belowbottomsep}
\end{tabular}
}

\label{tbl-SOTA-PC-IQA}
\end{table*}

\subsubsection{Qualitative Analysis} \label{sec-qualitative-enal}

\par In this section, a qualitative analysis of the the proposed uncertainty-driven framework is presented, with a focus on cases where our model exhibited high uncertainty. More precisely, the pairs selected were included in the top 10\% using the EIC criteria. These cases were carefully chosen to represent a spectrum of scenarios, including cases where the images of a pair have very different or similar perceptual qualities.

\par
\begin{figure*}[htbp]
\centerline{
\begin{tikzpicture}
    \node[inner sep=0pt] (image1) at (0,0) {
        \begin{tabular}{@{}c@{}}
            \includegraphics[scale=0.38]{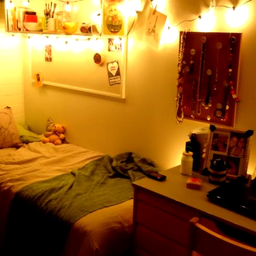} \\
            \small $Ref$ \\
        \end{tabular}
    };

    \node[inner sep=0pt] (image2) at (3.5,0) {
        \begin{tabular}{@{}c@{}}
            \includegraphics[scale=0.38]{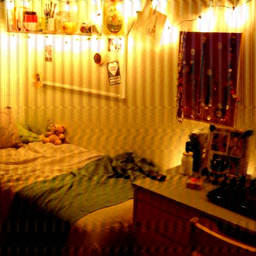} \\
            \small $A_1$ \\
        \end{tabular}
    };

    \node[inner sep=0pt] (image3) at (7,0) {
        \begin{tabular}{@{}c@{}}
            \includegraphics[scale=0.38]{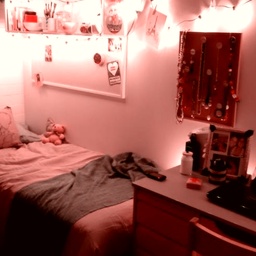} \\
            \small $B_1$ \\
        \end{tabular}
    };

    \node[inner sep=0pt] (image4) at (10.5,0) {
        \begin{tabular}{@{}c@{}}
            \includegraphics[scale=0.38]{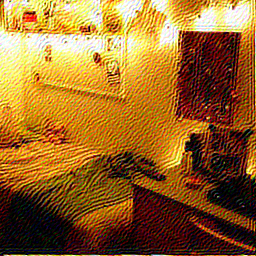} \\
            \small $A_2$ \\
        \end{tabular}
    };

    \node[inner sep=0pt] (image5) at (14,0) {
        \begin{tabular}{@{}c@{}}
            \includegraphics[scale=0.38]{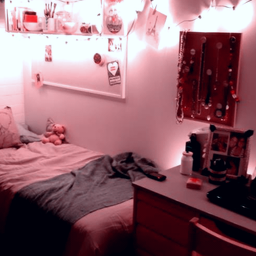} \\
            \small $B_2$ \\
        \end{tabular}
    };

    \draw[red, thick] (image2.south west) rectangle (image3.north east);
    \draw[blue, thick] (image4.south west) rectangle (image5.north east);

\end{tikzpicture}
}
\caption{Qualitative analysis for some selected pairs of a reference content. 
Red box: The first pair $(A_1, B_1)$ has the ground truth preference $Pr(A_1 \succ B_1) = 0.8$ and the predicted preference is $\hat{Pr}(A_1 \succ B_1) = 0.420$. Blue box: The second pair $A_2, B_2$ has the ground truth preference $Pr(A_2 \succ B_2) = 0.46$ and the predicted preference is $\hat{Pr}(A_2 \succ B_2) = 0.33$.}
\label{fig_first}
\end{figure*}

 Figure \ref{fig_first} depicts two illustrative cases. In the first instance (pairs $A_1$, $B_1$), where images have a clear difference in quality. Image $A_1$ was preferred over image $B_1$ with a ground truth probability of 0.8. However, the proposed model predicted a preference of 0.4, highlighting the challenge of capturing subjective human preferences. This is one of the cases where the pair should be selected for subjective assessment. Similarly, in the second case (pairs $A_2$, $B_2$), the decision is more difficult, the ground truth preference is around 0.46, while the proposed solution predicted 0.33. Thus, the decision is rather distortion preference than distortion visibility. In both of these cases, model uncertainty is high which reveals the difficulty of the proposed solution to estimate a robust probability of preference. 

\par
\begin{figure*}[htbp]
\centerline{
\begin{tikzpicture}
    \node[inner sep=0pt] (image1) at (0,0) {
        \begin{tabular}{@{}c@{}}
            \includegraphics[scale=0.38]{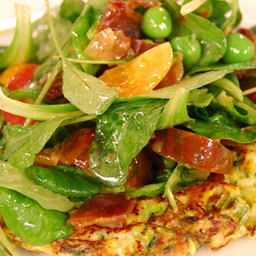} \\
            \small $Ref$
        \end{tabular}
    };

    \node[inner sep=0pt] (image2) at (3.5,0) {
        \begin{tabular}{@{}c@{}}
            \includegraphics[scale=0.38]{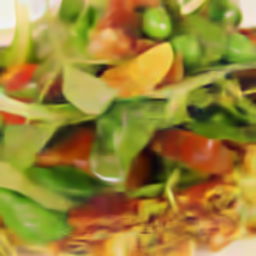} \\
            \small $A_1$
        \end{tabular}
    };

    \node[inner sep=0pt] (image3) at (7,0) {
        \begin{tabular}{@{}c@{}}
            \includegraphics[scale=0.38]{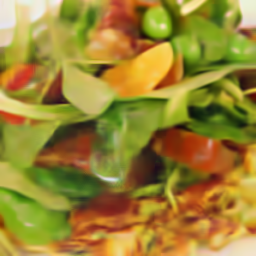} \\
            \small $B_1$
        \end{tabular}
    };

    \node[inner sep=0pt] (image4) at (10.5,0) {
        \begin{tabular}{@{}c@{}}
            \includegraphics[scale=0.38]{Figures/distort_183_spatiallyVaryingNoise_denoiser_2.png} \\
            \small $A_2$
        \end{tabular}
    };

    \node[inner sep=0pt] (image5) at (14,0) {
        \begin{tabular}{@{}c@{}}
            \includegraphics[scale=0.38]{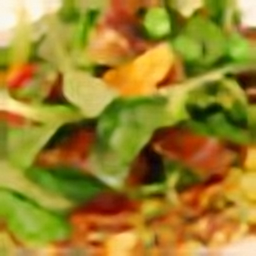} \\
            \small $B_2$
        \end{tabular}
    };

    \draw[red, thick] (image2.south west) rectangle (image3.north east);

    \draw[blue, thick] (image4.south west) rectangle (image5.north east);

\end{tikzpicture}
}
\caption{Qualitative analysis for some selected pairs of a reference content. 
Red box: The first pair $(A_1, B_1)$ has the ground truth preference $Pr(A_1 \succ B_1) = 0.35$ and the predicted preference is $\hat{Pr}(A \succ B) = 0.4$. Blue box: The second pair $A_2, B_2$ has the ground truth preference $Pr(A_2 \succ B_2) = 0.93$ and the predicted preference is $\hat{Pr}(A_2 \succ B_2) = 0.64$.}
\label{fig-qualitative}
\label{fig_second}
\end{figure*}

 To further investigate the uncertainty-based sampling method, two additional cases sourced from the same reference were examined, but in this case with a similar perceptual quality. In Fig. \ref{fig_second}, the red box, (pairs $A_1$, $B_1$), highlights a scenario where both images exhibit similar visual qualities with a ground truth preference of 0.35, and our model predicted a preference of 0.4. This demonstrates the method's sensitivity to subtle differences even when the distinction in quality is minimal. Moreover, in the pairs within the blue box (pairs $A_2$, $B_2$), although image $A_2$ was preferred over image $B_2$, the difference was not that distinguishable; the ground truth preference was 0.93, while the model predicted 0.64. This underscores the difficulty of aligning model predictions with human judgments when the visual differences are subtle.

\par The discrepancies between the model's predicted preferences and the ground truth preferences indicate areas where the model's performance can be improved. The model shows deviation in its predictions compared to the ground truth, which is measured by the expected information change (EIC). Higher EIC values suggest that by selecting this pair, the model's performance could be significantly enhanced. Overall, data and model uncertainty are the key factors that can guide the pairwise selection process. By selecting pairs for subjective evaluation the model's accuracy can be improved allowing better accuracy in the scores with a minimum cost.

\section{Conclusions and Future Work}\label{sec-conclusion}
In conclusion, the uncertainty-based sampling methods introduced in this work represent a significant improvement for pairwise comparison subjective assessment tests by reducing their duration while maintaining accurate quality scores. The LBPS-EIC framework stands out as a practical solution applicable to various image processing tasks, ensuring high-quality assessments with less human effort and lower computational costs. Utilizing deep learning models to predict human preferences and guide the sampling of pairs requiring human evaluation has proven highly effective. This approach models both aleatoric and epistemic uncertainty to achieve accurate preference predictions and optimize pairwise sampling, outperforming traditional active sampling methods. These advancements mark a substantial step forward in balancing the efficiency and precision of subjective quality assessments, ultimately benefiting both benchmarking and the training of learning-based quality metrics. 

\par For future work, it is planned to integrate both learning-based and active sampling approaches with a reinforcement learning framework, allowing an intelligent agent to learn to select pairs through interaction with human subjects. The use of vision transformers is also a promising direction for future work.  Transformer models and techniques such as hybrid CNN-transformer architectures have proven to enhance the quality prediction capabilities of the proposed framework. More importantly, attention maps computed in the transformer model across multiple heads and layers could be exploited to evaluate the model confidence in its prediction.

\par As a final remark, the uncertainty-based sampling methods presented in this work represent a new way to perform subjective assessment, where not only human subjects are used but also objective quality metrics. This means that humans and machines cooperate together for the same goal, allowing us to annotate large-scale datasets of visual data.

\section*{Acknowledgments}
This work was supported by the Fundação para a Ciência e a Tecnologia (FCT), Portugal, through the Project entitled Deep Compression: Emerging Paradigm for Image Coding under Grant PTDC/EEI-COM/7775/2020.

\bibliographystyle{IEEEtran}
\bibliography{reference}


\begin{IEEEbiography}[{\includegraphics[width=1in,height=1.25in,clip,keepaspectratio]{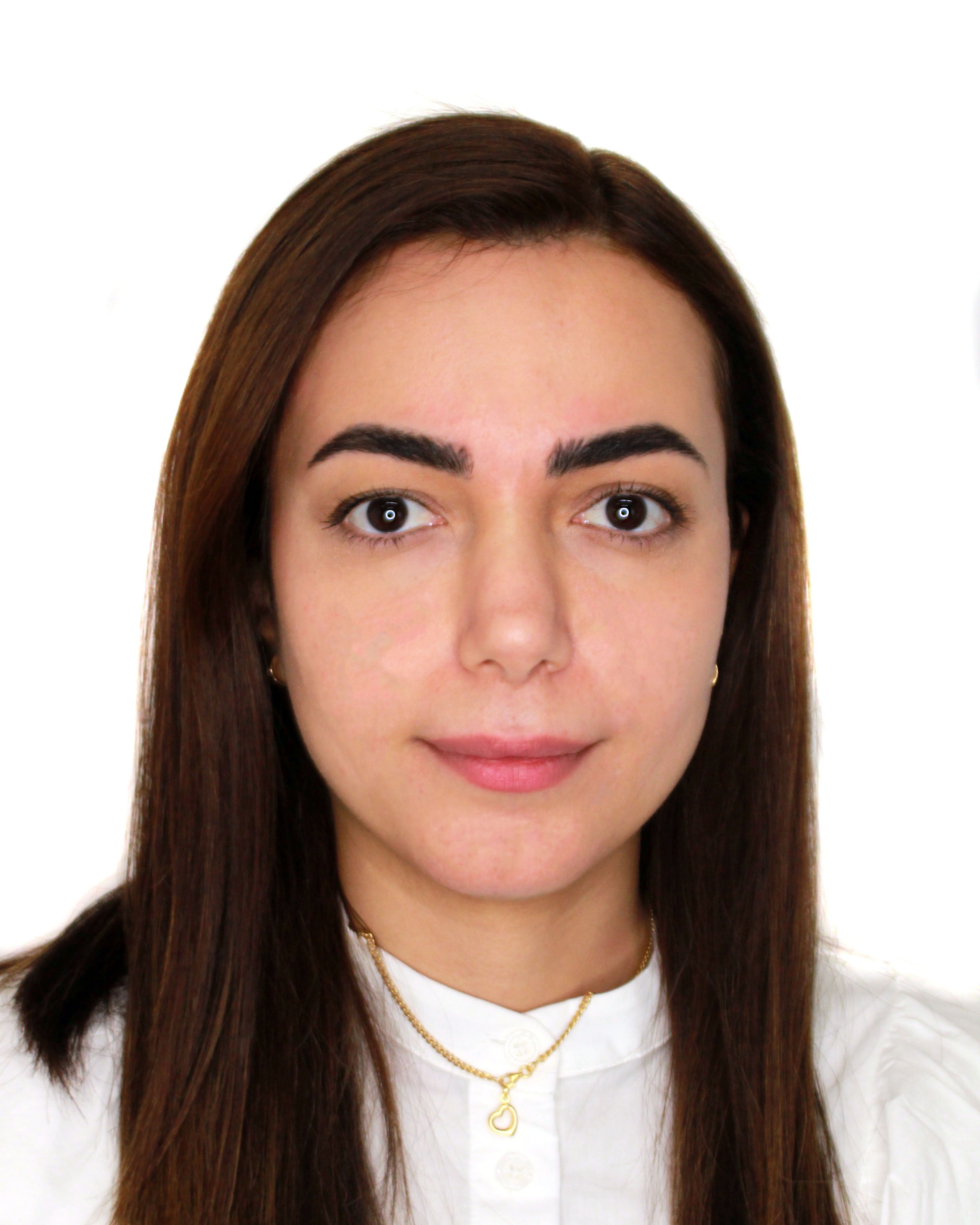}}]{Shima Mohammadi} (Graduate Student Member, IEEE) 
received the M.Sc. degree in Electrical and Computer Engineering from the University of Tehran, Tehran, Iran, in 2020. She is currently a Ph.D. student with the Department of Electrical and Computer Engineering, Instituto Superior Técnico (IST), Universidade Técnica de Lisboa, Lisbon, Portugal, and a member of the Instituto de Telecomunicações. Her research interests include visual coding, quality assessment, and machine learning applications.
\end{IEEEbiography}

\begin{IEEEbiography}[{\includegraphics[width=1in,height=1.25in,clip,keepaspectratio]{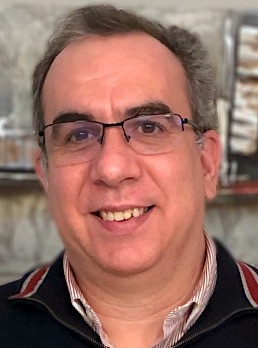}}]{Joao Ascenso} (Senior Member, IEEE)
received the E.E., M.Sc., and Ph.D. degrees in electrical and computer engineering from the Instituto Superior Técnico (IST), Universidade Técnica de Lisboa, Lisbon, Portugal, in 1999, 2003, and 2010, respectively. He is currently an Associate Professor with the Department of Electrical and Computer Engineering, IST, and a member of the Instituto de Telecomunicações. He has published more than 150 papers in international conferences and journals. His current research interests include visual coding, quality assessment, coding and processing of 3D visual representations, coding for machines, super-resolution, denoising among others. He was an associate editor of IEEE Transactions on Image Processing, IEEE Signal Processing Letters and IEEE Transactions on Multimedia and guest editor of IEEE Access and IEEE Transactions on Circuits and Systems for Video Technology. He received three Best Paper Awards at PCS 2015, ICME 2020, and MMSP 2024 and has served as Technical Program Chair and in other organizing committees of major international conferences, including IEEE ICIP, PCS, EUVIP, ICME, MMSP and ISM. 
\end{IEEEbiography}

\vfill
\end{document}